\documentclass[twocolumn,pre,preprintnumbers,superscriptaddress,notitlepage]{revtex4}
\usepackage{textcomp}
\usepackage{amsmath}
\usepackage{amssymb}
\usepackage{graphicx,graphics,dcolumn,bm,enumerate}
\usepackage{color}
\usepackage{soul}
\usepackage{array}
\usepackage{comment,natbib,appendix}
\usepackage{lipsum}
\usepackage{epsfig}
\usepackage{epstopdf}
 \usepackage{pgfplots}
 \usepackage{hyperref}

\newcommand{\be}{\begin{equation}}
\newcommand{\ee}{\end{equation}}
\newcommand{\ba}{\begin{eqnarray}}
\newcommand{\ea}{\end{eqnarray}}

\newcommand{\cmp}
{\affiliation{Condensed Matter Physics Division, 
Saha Institute of Nuclear Physics, 1/AF Bidhannagar, Kolkata 700064, India.}}
\newcommand{\barasat}
{\affiliation{Barasat Government College, Barasat, Kolkata 700124, India.}}

\begin{document}

\title{Search of the ground state(s) of spin glasses and quantum annealing}

 \author{Sudip Mukherjee}
 \email{sudip.mukherjee@saha.ac.in}
 \barasat \cmp

 \begin{abstract}
We review our earlier studies on the order parameter distribution of the 
quantum Sherrington-Kirkpatrick (SK) model. Through Monte Carlo technique, 
we investigate the behavior of the order parameter distribution at finite 
temperatures. The zero temperature study of the spin glass order parameter 
distribution is made by the exact diagonalization method. We find in low-temperature 
(high-transverse-field) spin glass region, the tail (extended up to zero value 
of order parameter) and width of the order parameter distribution become zero 
in thermodynamic limit. Such observations clearly suggest the existence of a 
low-temperature (high-transverse-field) ergodic region. We also find in 
high-temperature (low-transverse-field) spin glass phase the order parameter 
distribution has nonzero value for all values of the order parameter even in 
infinite system size limit, which essentially indicates the nonergodic behavior 
of the system. We study the annealing dynamics by the paths which pass through 
both ergodic and nonergodic spin glass regions. We find the average annealing 
time becomes system size independent for the paths which pass through the 
quantum-fluctuation-dominated ergodic spin glass region. In contrast to that, 
the annealing time becomes strongly system size dependent for annealing down 
through the classical-fluctuation-dominated nonergodic spin glass region. We 
investigate the behavior of the spin autocorrelation in the spin glass phase. 
We observe that the decay rate of autocorrelation towards its equilibrium value 
is much faster in the ergodic region with respect to the nonergodic region of 
the spin glass phase.    
\end{abstract}
\maketitle

\section{Introduction}
Many-body localization has been a topic of major interest and research, following the
classic work of Anderson (see e.g.,~\cite{sudip-Nandkishore} for a recent review).
Ray et al.~\cite{sudip-ray} published a paper in 1989, which
shows the possibility of the delocalization in the Sherrington-Kirkpatrick (SK)~\cite{sudip-sk} spin glass system 
with the help of quantum fluctuation. The free energy landscape of SK spin glass is highly rugged. The free energy 
barriers are macroscopically high, which separate the local free energy minima. The system often get trapped into 
any one of such local free energy minima and as a result of that one would find a board spin glass order parameter 
distribution, which contains a long tail extended up to the zero value of the order parameter as suggested by 
Parisi~\cite{sudip-parisi1,sudip-parisi2}. Such broad order parameter distribution indicates breaking of replica 
symmetry, which essentially reveals the nonergodic behavior of the system. This nonergodicity is responsible for the 
NP hardness in the search of the spin glass ground state(s) and equivalent optimization problems (see e.g.,~\cite{sudip-das}). 

The situation can be remarkably different if SK spin glass is placed under the transverse field. Using the quantum 
fluctuation the system can tunnel through the high (but narrow) free energy barriers and consequently the system 
can avoid the trapping in the local free energy minima. Such phenomena of quantum tunneling across the free energy 
barriers was first reported by Ray et al.~\cite{sudip-ray}. This key idea plays instrumental role in the development 
of the quantum annealing~\cite{sudip_Kadowaki,sudip_brooke,sudip_farhi,sudip_santoro,sudip-das,sudip_Johnson,sudip-bikas,sudip-lidar}.  
With the aid of quantum fluctuation the system can explore the entire free energy landscape and essentially the system 
regains its ergodicity. Therefore, one would expect a narrow order parameter distribution, sharply peaked about any 
nonzero value of the order parameter. 

We investigate the behavior of the spin glass order parameter distribution of the quantum SK spin glass~\cite{sudip-op_dis}. 
At finite temperature, employing Monte Carlo method, we numerically extract the spin glass order parameter distribution 
whereas for zero temperature we use exact diagonalization technique. From our numerical results we identify a low-temperature 
(high-transverse-field) spin glass region where the tail of the order parameter distribution vanishes in the thermodynamic limit. We 
observe in such quantum-fluctuation dominated region, the order parameter distribution shows clear tendency of becoming a 
delta function for infinite system size, which essentially indicates the ergodic nature of system. On the other hand, we find
in high-temperature (low-transverse field) spin glass phase, the order parameter distribution remains Parisi type~\cite{sudip-young}, 
suggesting the nonergodic behavior of the system. We perform dynamical study of the system to investigate the variation of 
the average annealing time in both ergodic and nonregodic spin glass regions. Using the effective Suzuki-Trotter Hamiltonian 
dynamics of the model, we try to reach a fixed low-temperature and low-transverse-field point along the annealing paths, which 
are passing through the both ergodic and nonergodic regions. With limited system sizes, we do not find any system size dependence  
of the annealing time when such dynamics is performed along the paths going through the quantum-fluctuation dominated ergodic 
spin glass region. In case of annealing down through the nonregodic region, we clearly observe the increase in the annealing 
time with increase of system size. 

We examine the nature of spin autocorrelation in the spin glass phase of the quantum SK model~\cite{sudip-jpn}. We find the 
relaxation behavior of the spin autocorrelation is very different in ergodic and nonergodic regions. In quantum-fluctuation-dominated  
ergodic spin glass phase, the autocorrelation relaxes extremely quickly whereas the effective relaxation time is much higher in 
the classical-fluctuation-dominated nonregodic spin glass region.

\section{Models}
The Hamiltonian of the quantum SK model containing $N$ Ising spins is given by (see e.g.,~\cite{sudip-ttc-book})
\begin{align}
H  = H_0 + H_I;~ H_0=-\sum_{i < j} J_{ij}\sigma_i^z\sigma_j^z;~ H_I=-{\Gamma} \sum_{i = 1}^N\sigma_i^x . \label{Ham}  
\end{align}
Here $\sigma_i^z$ and $\sigma_i^x$ are the $z$ and $x$ components of Pauli spin matrices respectively. The spin-spin 
interactions $J_{ij}$ are distributed following a Gaussian distribution 
$\rho (J_{ij}) = \Big (\frac{N}{2{\pi}J^2}\Big)^{\frac{1}{2}}\exp\Big (\frac{-NJ_{ij}^2}{2J}\Big)$ with zero mean and 
standard deviation $J/\sqrt{N}$. The transverse field is denoted by $\Gamma$. To perform finite temperature ($T$) study 
on the quantum SK spin glass, the Hamiltonian $H$ is mapped into an effective classical Hamiltonian $H_{eff}$ 
using Suzuki-Trotter formalism. Such $H_{eff}$ is given by \begin{align}
 H_{eff}=-\sum_{n=1}^M \sum_{i < j} {J_{ij}\over M}\sigma_i^n\sigma_j^n-\sum_{i=1}^N\sum_{n=1}^M{1\over {2\beta}}\text{log~coth}{\beta\Gamma\over M}\sigma_i^n\sigma_i^{n+1} \label{H_cl}. 
\end{align}
Here $\beta$ is the inverse of temperature and $\sigma_i^n=\pm 1$ is the classical Ising spin. One can see the appearance 
of an additional dimension in Eq.~(\ref{H_cl}), which is often called Trotter dimension $M$. As $\beta \to \infty$ the $M \to \infty$.

\begin{figure}[htb]
\begin{center}
\includegraphics[width=7.0cm]{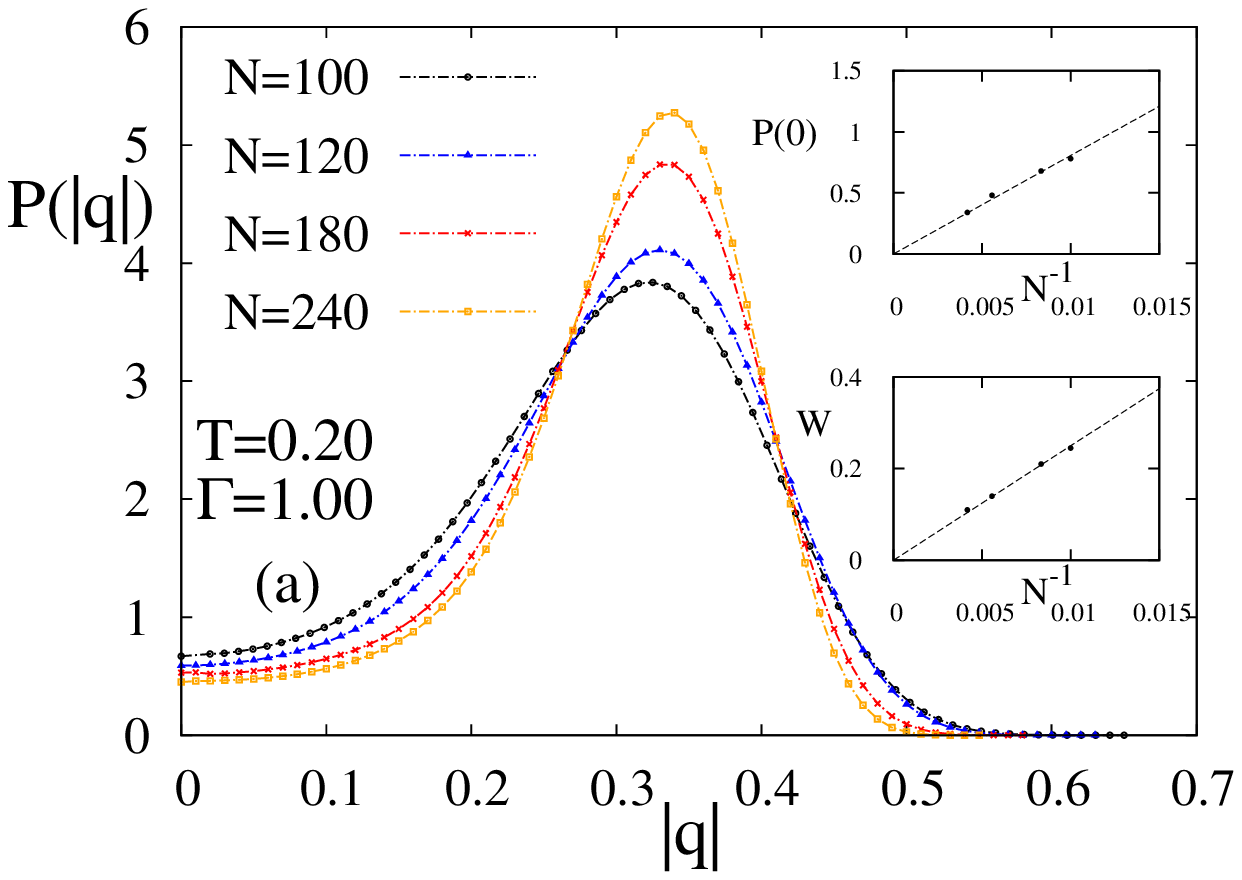}
\includegraphics[width=7.0cm]{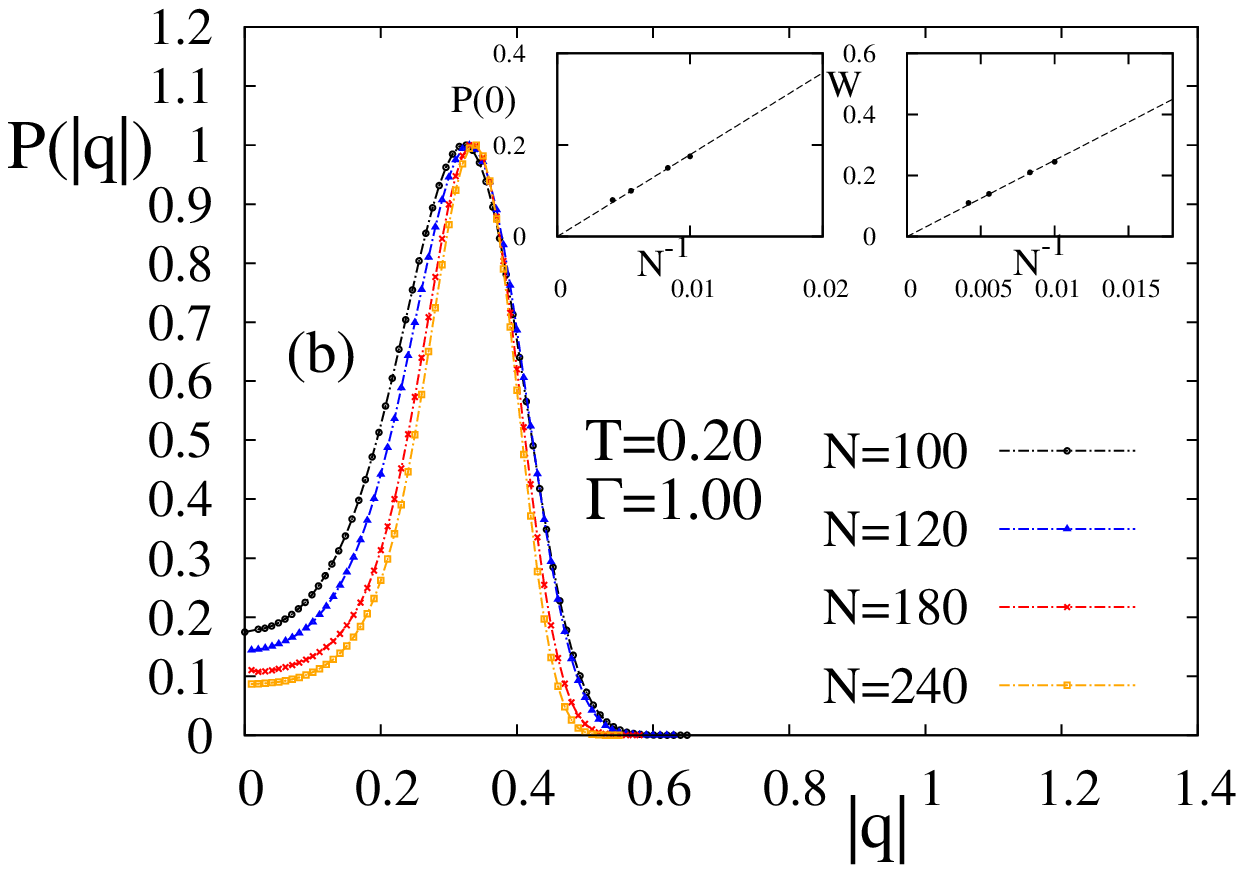}
\end{center}
\caption{For temperature $T = 0.20$ and  transverse field $\Gamma = 1.00$, the plots of the order parameter 
distribution $P(|q|)$: (a) area-normalized (b) peak-normalized. The numerical data are obtained from Monte 
Carlo simulations. Extrapolations of $P(0)$ and $W$ (width of the distribution function) with $1/N$ are shown in
the insets. For both area and peak normalized distributions, the $P(0)$ and $W$ tend to zero for infinite system size.}
\label{ergodic_figs}
\end{figure}

\begin{figure}[htb]
\begin{center}
\includegraphics[width=5.5cm]{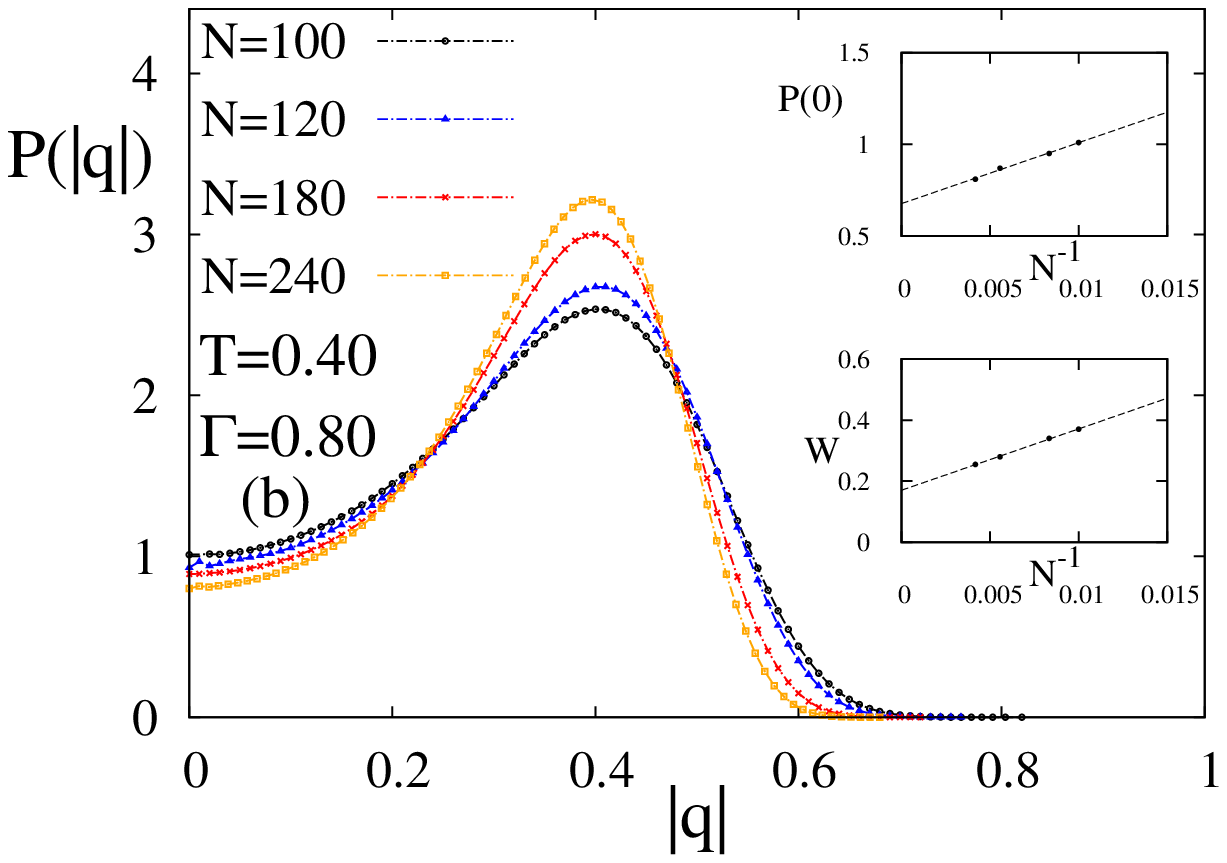}
\includegraphics[width=5.5cm]{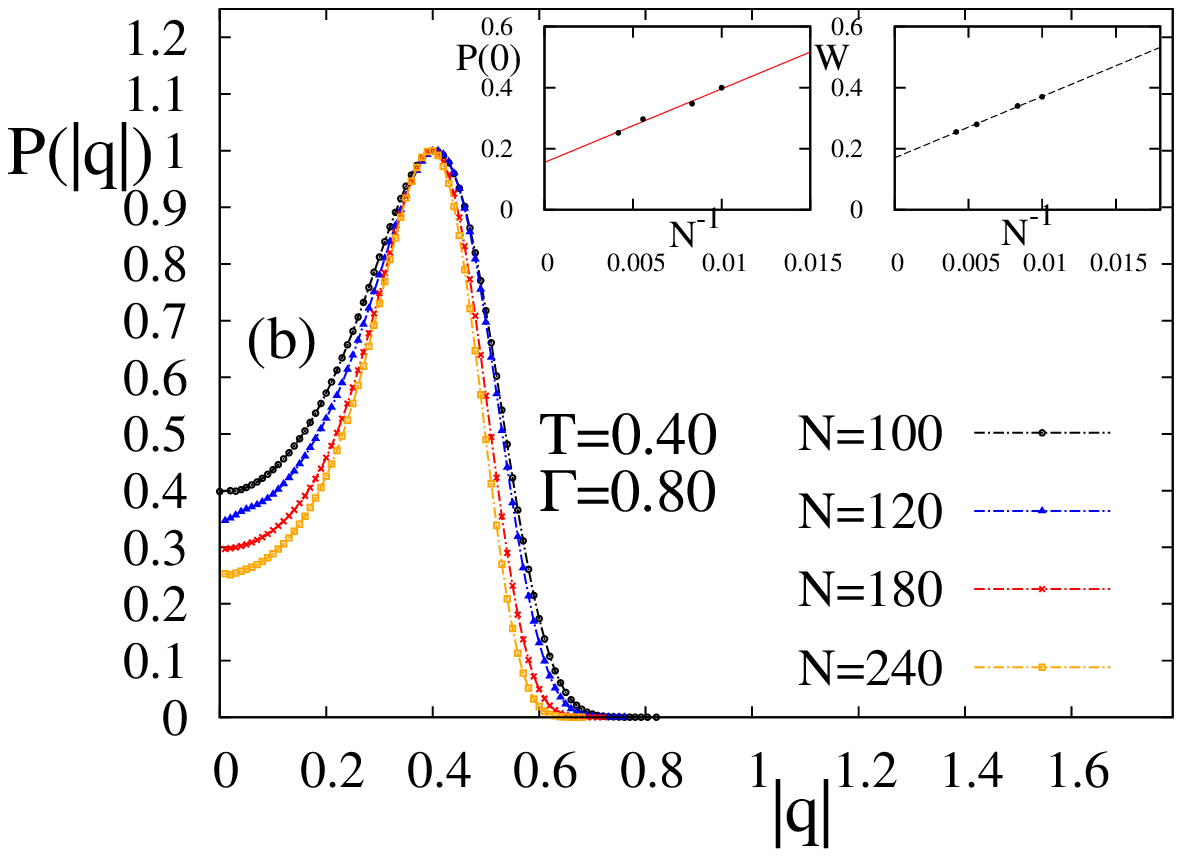}
\end{center}
\caption{For temperature $T = 0.40$ and  transverse field $\Gamma = 0.80$, the plots of the order parameter 
distribution $P(|q|)$: (a) area-normalized (b) peak-normalized. The numerical data are obtained from Monte 
Carlo simulations. Extrapolations of $P(0)$ and $W$ (width of the distribution function) with $1/N$ are shown in
the insets. For both area and peak normalized distributions, the $P(0)$ and $W$ remain finite even in thermodynamic limit.}
\label{nonergodic_figs}
\end{figure}

\section{Finite temperature study of order parameter distribution in the spin glass phase}
We perform the Monte Carlo simulation on the $H_{eff}$ to extract the behavior of the order parameter distribution at 
finite temperature~\cite{sudip-op_dis}. We first allow the system for equilibration with $t_0$ Monte Carlo steps and 
then at each Monte Carlo step we compute the replica overlap 
$q^{{\alpha}{\beta}}(t) = \frac{1}{NM}\sum_{i=1}^N\sum_{m=1}^M(\sigma_i^m(t))^{\alpha}(\sigma_i^m(t))^{\beta}$, where 
$(\sigma_i^m)^{\alpha}$ and $(\sigma_i^m)^{\beta}$ denote the spins of the two identical replicas $\alpha$ and $\beta$ respectively,  
having same set of spin-spin interactions. 
We define one Monte Carlo step as a sweep over the entire system, where each spin is updated once. The order parameter 
distribution is given by 
\begin{align*}
 P(q)=\overline{ \frac{1}{t_1}\sum_{t=t_0}^{t_0+t_1}\delta(q-q^{\alpha \beta}(t)) } .
\end{align*}
Here overhead bar denotes the averaging over the configuration. For thermal averaging we consider $t_1$ Monte Carlo 
steps. The order parameter of the spin glass system is define as 
$q = \frac{1}{MN}\sum_{m=1}^{M}\sum_{i=1}^{N}\overline{{\langle \sigma_i^m \rangle}^2}$, where $\langle .. \rangle$ 
indicates the thermal average for given set disorder. We numerically obtain the order parameter distribution $P(q)$ 
for a given set of $T$ and $\Gamma$. Along with the usual area normalized distribution, we also evaluate the peak 
normalized order parameter distribution (where the peak is normalized by its maximum value). 

To perform Monte Carlo simulations, we take system sizes $N  = 100, 120, 180, 240$ and number of Trotter slices $M = 15$. 
The equilibrium time of the system is not identical throughout the entire spin glass region on $\Gamma - T$ plane. 
We find the equilibrium time of the system (for $100 \leq N \leq 240$) is typically $\lesssim 10^6$ within the region 
$T < 0.25$ and $\Gamma < 0.40$, whereas for the rest of the spin glass region the systems equilibrates within $\leq 10^5$ 
Monte Carlo steps. In our numerical simulations, we take $J = 1$ and the thermal averaging is made over the $t_1 = 1.5 \times 10^5$ 
time steps. The configuration average is made over $1000$ sets of realizations. Due to the presence of ${\mathbb{Z}}_2$ 
symmetry in the system, we evaluate the distribution of $|q|$ instead of $q$. We find a clear system size dependence of 
$P(0)$ and we extrapolate $P(0)$ with $1/N$ to find its behavior in thermodynamic limit. We also evaluate the width $W$ 
of the distribution which is define as $W = |q_2 - q_1|$. The value of the distribution becomes half of its maximum at 
$q= q_1$ and $q_2$. Like $P(0)$, we also extrapolate $W$ with $1/N$ to get the its nature in infinite system size limit. 
We observe two distinct behaviors of the extrapolated values of $P(0)$ and $W$ in the two different regions of spin 
glass phase. In low-temperature (high-transverse field) region of the spin glass phase, both $P(0)$ and $W$ go to zero in 
infinite system size limit [see Fig.~\ref{ergodic_figs}(a)]. The asymptotic behaviors of $P(0)$ and $W$ remain same for the 
peak normalized order parameter distribution in this region of the spin glass phase [see Fig.~\ref{ergodic_figs}(b)]. Therefore 
in low-temperature (high transverse) spin glass region, the behaviors of $P(0)$ and $W$ indicate the order parameter 
distribution would eventually approaches to Gaussian form in thermodynamic limit, which suggests the ergodic nature of 
the system. One the other hand, in high-temperature (low-transverse field) spin glass region, we find neither $P(0)$ nor 
$W$ goes to zero even in thermodynamic limit [see Fig.~\ref{nonergodic_figs}(a)]. Again, for peak normalized distribution in 
low-temperature (high-transverse field) spin 
glass region, the extrapolated values of $P(0)$ and $W$ remain finite in large system size limit [see Fig.~\ref{nonergodic_figs}(b)].  
That means in high-temperature (low-transverse field) spin glass region, the order parameter distribution has no tendency 
to take the Gaussian form and it contains long tail (extended to zero value of $q$) even in thermodynamic limit. This indicates 
the nonregodic behavior of the system in the high-temperature (low transverse field) spin glass region. 

\begin{figure}[ht]
\begin{center}
\includegraphics[width=5.5cm]{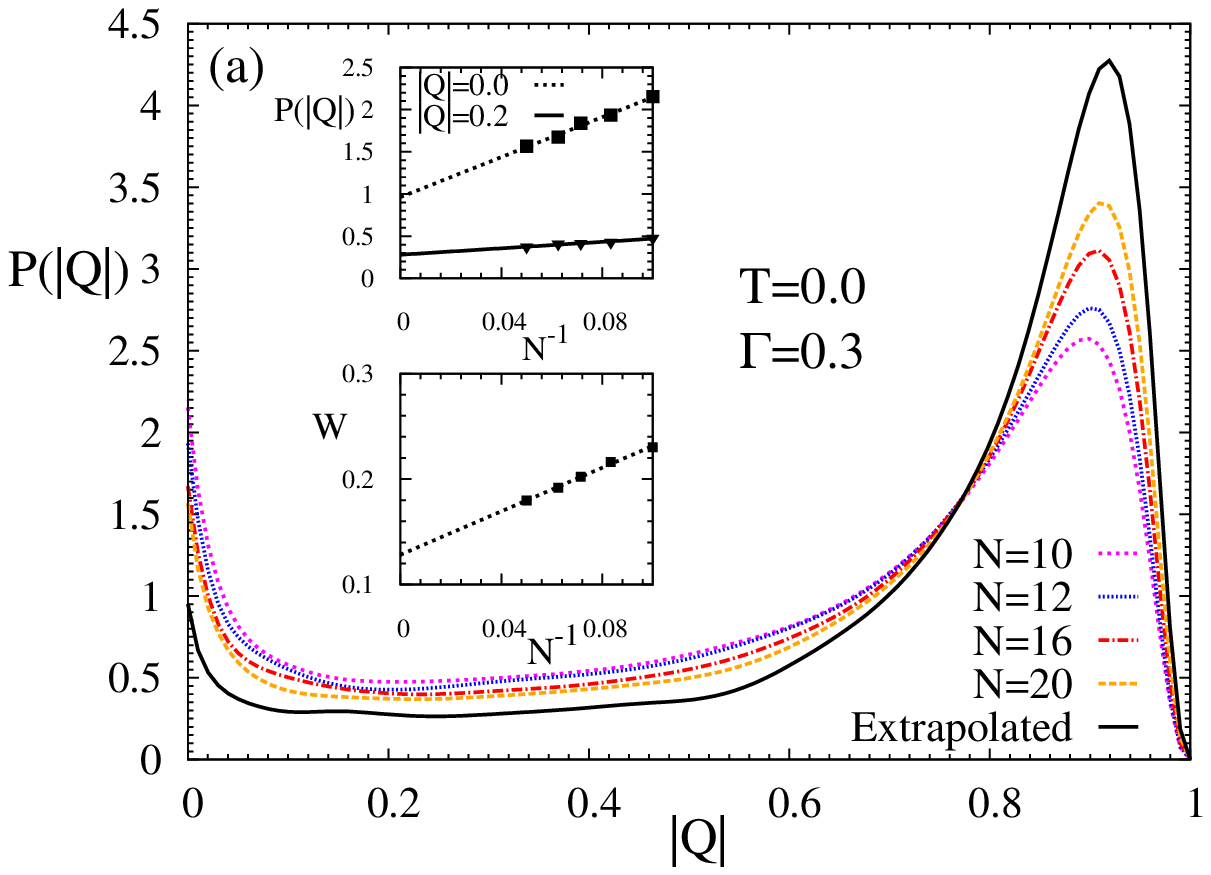}
\includegraphics[width=5.5cm]{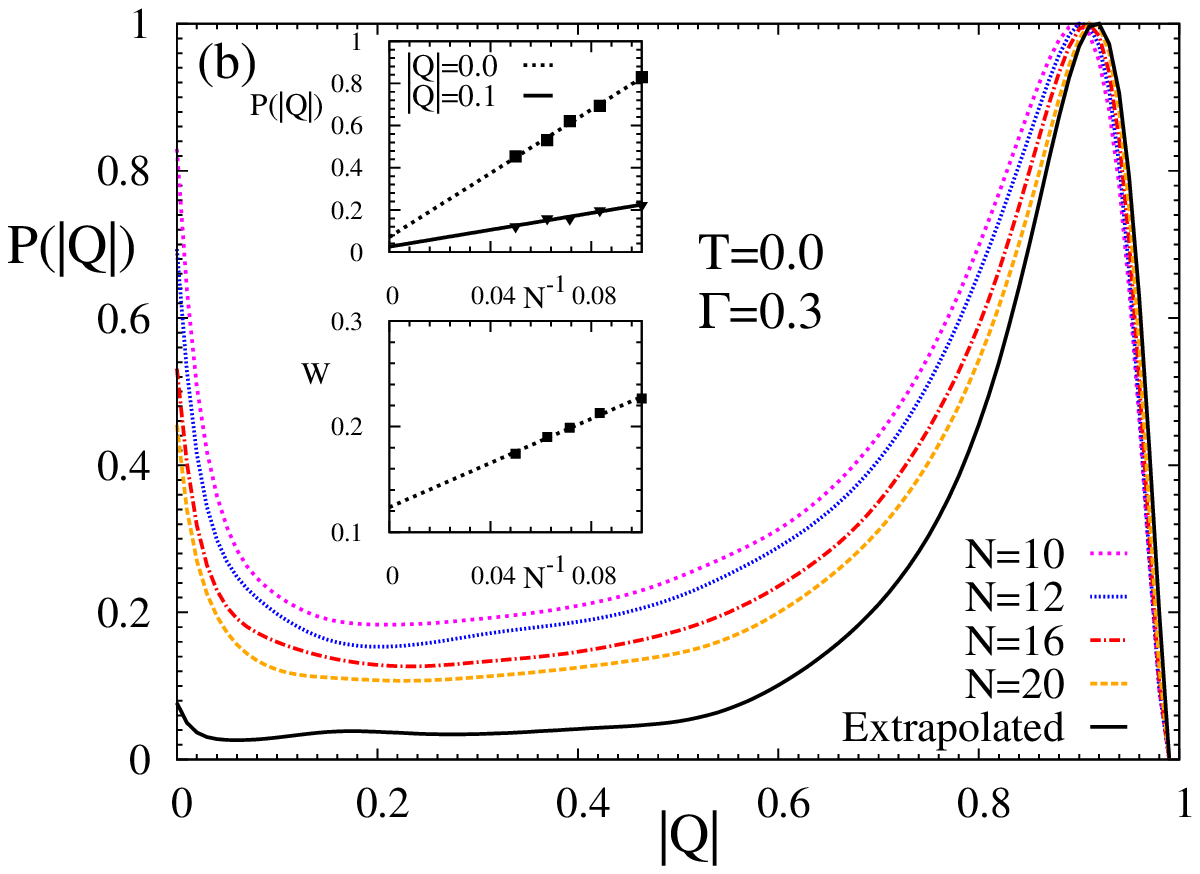}
\end{center}
\caption{Plots of $P(|Q|)$ at $T = 0$ and $\Gamma = 0.30$ with four different system, obtained from exact 
diagonalization method. For (a) the area under the each $P(|Q|)$ curve is normalized to unity; (b) the peaks 
of the all $P(|Q|)$ curves are normalized by their maximum values. For area-normalized distributions, typical 
extrapolations of $P(|Q|)$ with $1/N$ for  $|Q|=0.0$ and $0.2$ are shown in the top inset, whereas for 
peak-normalized distributions such extrapolations are shown in the top inset for $|Q|=0.0$ and $0.1$. In each 
figure, the bottom inset shows the extrapolation of $W$ as a function of $1/N$.  
 }
\label{op_dist_0.3gama}
\end{figure}

\section{Zero temperature study of order parameter distribution in the spin glass phase}
Using exact diagonalization technique we explore the behavior of the spin glass order parameter distribution at zero temperature. 
We obtain the ground state of the Hamiltonian in Eq.~(\ref{Ham}) through Lanczos algorithm. We express the $H$ in the spin basis 
states, which are actually the eigenstates of $H_0$. We obtain (after diagonalization) the eigenstates of $H$, where $n$-th 
eigenstate can be written as $|\psi_n\rangle~= \sum_{\alpha=0}^{2^{N-1}} a_{\alpha}^n |\varphi_\alpha\rangle$. Here $|\varphi_\alpha\rangle$ 
are the eigenstates of $H_0$ and  $a_{\alpha}^n=\langle\varphi_{\alpha}|\psi_n\rangle$. Since in the present case we confine 
our study at zero temperature, then only the ground state averaging is made in the evaluation of order parameter. The order parameter at 
zero temperature is define as $Q = (1/N) \sum_i \overline{\langle\psi_0|\sigma_i^z|\psi_0\rangle^2}=(1/N)\sum_i\overline{Q_i}$, 
where $Q_i$ is the local site-dependent order parameter. One can see that the definition of the order parameter at zero temperature 
is different form its definition at the finite temperature. The oder parameter distribution is given by 
\begin{equation}
P(|Q|)=\overline{\frac{1}{N}\sum_{i=1}^N\delta(|Q|-Q_i)}. \nonumber
\end{equation}
We study the behavior of order parameter distribution (at $T = 0$) with the system sizes $N  = 10, 12, 16, 20$. Like the finite 
temperature, for a given value of $\Gamma$ we numerically obtain both area [see the Fig.~\ref{op_dist_0.3gama}(a)] and peak 
[see the Fig.~\ref{op_dist_0.3gama}(b)] normalized order parameter distributions. In both the cases one can observe that,  
beside a peak at any nonzero value of $|Q|$, $P(|Q|)$ also shows an upward rise for low values of $|Q|$. However, the value
of $P(0)$ decrease with increase in the system size. To obtain the behavior of both area and peak normalized $P(|Q|)$ in 
thermodynamic limit,  we extrapolate the $P(|Q|)$ with $1/N$ for each values of $|Q|$. For $\Gamma = 0.30$, the extrapolations 
of area normalized $P(|Q|)$ at $|Q| = 0.0, 0.2$ are shown in the top inset of Fig.~\ref{op_dist_0.3gama}(a). Similarly the 
extrapolations of peak normalized $P(|Q|)$ at $|Q| = 0.0, 0.1$ are shown in the top inset of Fig.~\ref{op_dist_0.3gama}(b) for 
the same value of $\Gamma$. In addition to the extrapolation of $P(|Q|)$, we also extrapolate the $W$ with $1/N$ where $W$ is 
the width of the order parameter distribution. The $W$ is define as $W = |Q_2 - Q_1|$, where $P(|Q|)$ becomes half of its maximum 
at the values $|Q_1|$ and $|Q_2|$. The extrapolations of $W$ as a function of $1/N$ for area and peak normalized distributions 
are shown in the bottom insets of Fig.~\ref{op_dist_0.3gama}(a) and Fig.~\ref{op_dist_0.3gama}(b) respectively. Due to the severe 
limitation of maximum system, the extrapolated curve of $P(|Q|)$ (for infinite system size) does not take the form of delta function    
even in thermodynamic limit. However, the order parameter distribution shows clear tendency of getting narrower with the increase 
of system size. The limitation in system size, is also responsible for the nonzero value of $W$ even in infinite system size limit.  
Hence, we can say that at zero temperature, in the spin glass phase, the $P(|Q|)$ would eventually become a delta function (peaked 
about a finite value of $|Q|$) in thermodynamic limit, which essentially suggests the ergodic behavior of the system.   

\begin{figure}[h]
\begin{center}
\includegraphics[width=6.3cm]{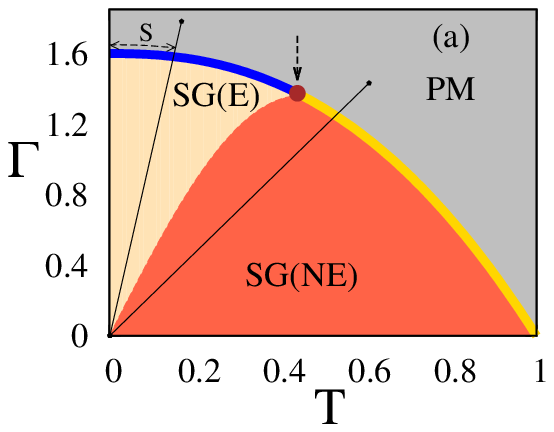}
\includegraphics[width=6.8cm]{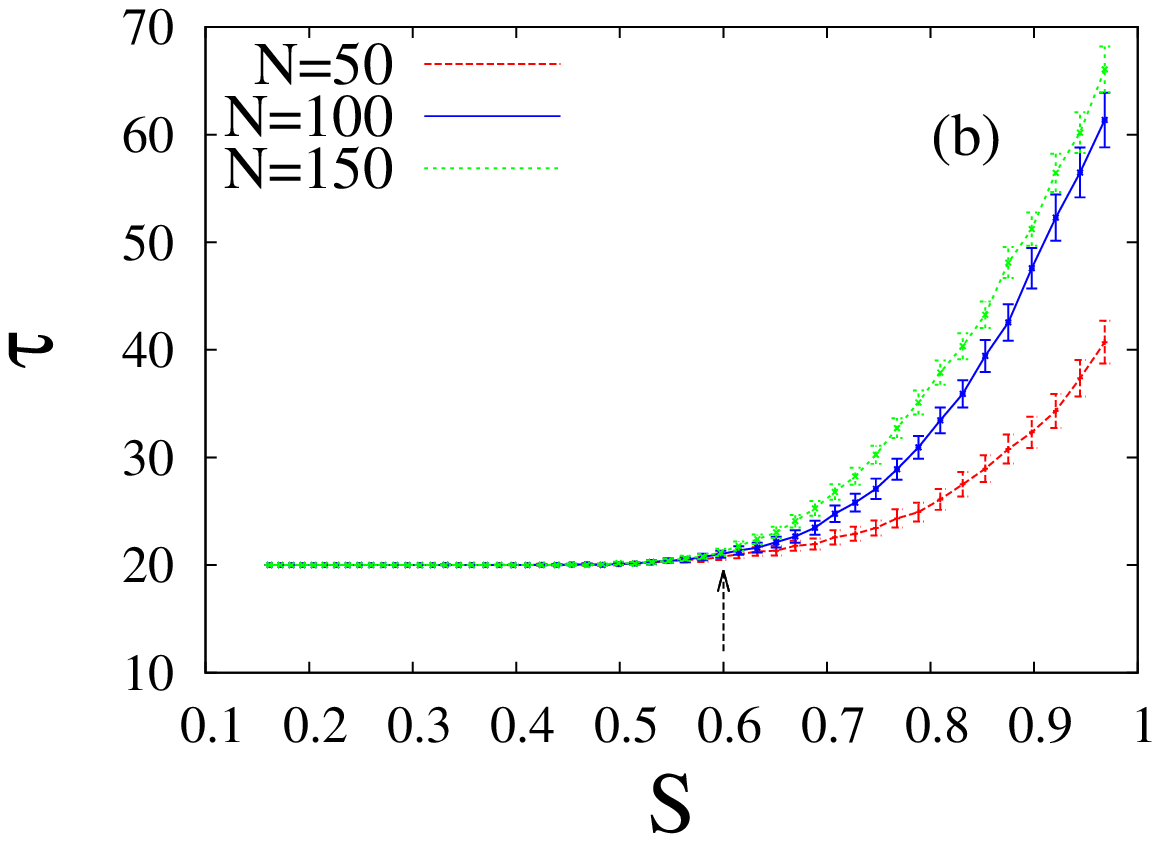}
\end{center}
\caption{  
(a) Schematic phase diagram of the quantum SK model \cite{sudip-cl_qm}. 
The PM and SG represent paramagnetic and spin glass phases respectively.
In context of ergodicity, the spin glass phase is further divided into 
two regions: ergodic spin glass region SG(E) and nonergodic spin glass SG(NE). 
The quantum-classical crossover point\cite{sudip-cl_qm,sudip-yao} in 
the critical behavior of the SK model is shown by the filled-red circle on 
the SG-PM phase boundary. We accomplish annealing following the linear paths, 
passing through both SG(E) and SG(NE) regions. Among such annealing paths, 
two of them are indicated by the inclined straight lines.
(b) Variation of annealing time $\tau$ with arc-length $S$ (cf.~\cite{sudip-op_dis}).
Such arc-length is actually the distance measured along the phase boundary, starting 
from zero-temperature quantum critical point ($T=0, \Gamma \simeq 1.6$), up to the 
intersection point of the phase boundary with the annealing path. The error bars indicate 
the errors associated with the numerical data.  Up to the arc-length distance 
$S = 0.60 \pm 0.05$ (corresponds to $T = 0.49  \pm 0.03 , \Gamma = 1.31 \pm 0.04$), indicated 
by vertical arrows in both figures, the annealing time fairly remains system size independent. 
Whereas $\tau$ increases rapidly with the system size when the annealing paths pass through 
the SG(NE) region.
 }
\label{EE_NEE_phase_diagram}
\end{figure}

\section{Annealing through ergodic and nonergodic regions}
In the previous sections we find, in the low-temperature (high-transverse field) spin glass phase the order parameter 
distribution becomes delta function in thermodynamic limit. Such feature suggests the ergodic behavior of the system 
in the low-temperature (high-transverse field) spin glass region. In contrast, we also observe in high-temperature 
(low-transverse field) spin glass region the order parameter distribution remains Parisi type, indicating the 
nonergodic behavior of the system. These ergodic and nonergodic regions are separated by the line originates from 
$T = 0$, $\Gamma = 0$ and touches the spin glass phase boundary at the quantum-classical crossover point~\cite{sudip-cl_qm,sudip-yao}. 
To study the dynamical behaviors of these regions, we perform annealing using the $H_{eff}$ with time dependent $T$ 
and $\Gamma$~\cite{sudip-op_dis}. We consider linear annealing schedules; $T(t) = T_0(1 - \frac{t}{\tau})$ and 
${\Gamma}(t) = {\Gamma}_0(1 - \frac{t}{\tau})$. Here $T_0$ and ${\Gamma}_0$ belong to the paramagnetic region and they 
are equidistant from the phase boundary in the different parts of the phase diagram. We should mention that, to avoid 
the singularities in the $H_{eff}$, we are forced to keep very small values ($\simeq 10^{-3}$) of both $T$ and $\Gamma$ 
even at the end of the annealing schedules. The annealing time is denoted by $\tau$, which is actually the time to achieve 
a very low free energy corresponding to the very small values of $T \simeq 10^{-3} \simeq \Gamma$. We explore the annealing 
dynamics of the system for a path that either passes through the ergodic or nonergodic spin glass regions 
[see Fig~\ref{EE_NEE_phase_diagram}(a)]. We study the 
variation of $\tau$ with $S$, which is define as the arc-distance between the intersection point of the annealing path with 
the critical line and the zero temperature spin glass to paramagnetic transition point. Such arc-distance is measured 
along the phase boundary between the spin glass and paramagnetic phases. Our numerical results show that upto a certain value 
of arc-length [$S \lesssim 1$; see Fig~\ref{EE_NEE_phase_diagram}(b)], the annealing time does not have any system size dependence. 
These results actually associated with paths which are passing through the ergodic region SG(E) of the spin glass phase. 
On the other hand, for $S \gtrsim 1$, corresponding to the paths passing through the nonergodic spin glass region, the $\tau$ 
increases with the increase of $S$. We also find that, the numerical error associated with the estimation of $\tau$, 
increases monotonically with $S$ and beyond $S \gtrsim 1$ such error bars in $\tau$ corresponding to different values of 
$N$ start overlapping  [see Fig~\ref{EE_NEE_phase_diagram}(b)].

\begin{figure*}[htb]
\begin{center}
\includegraphics[width=6.0cm]{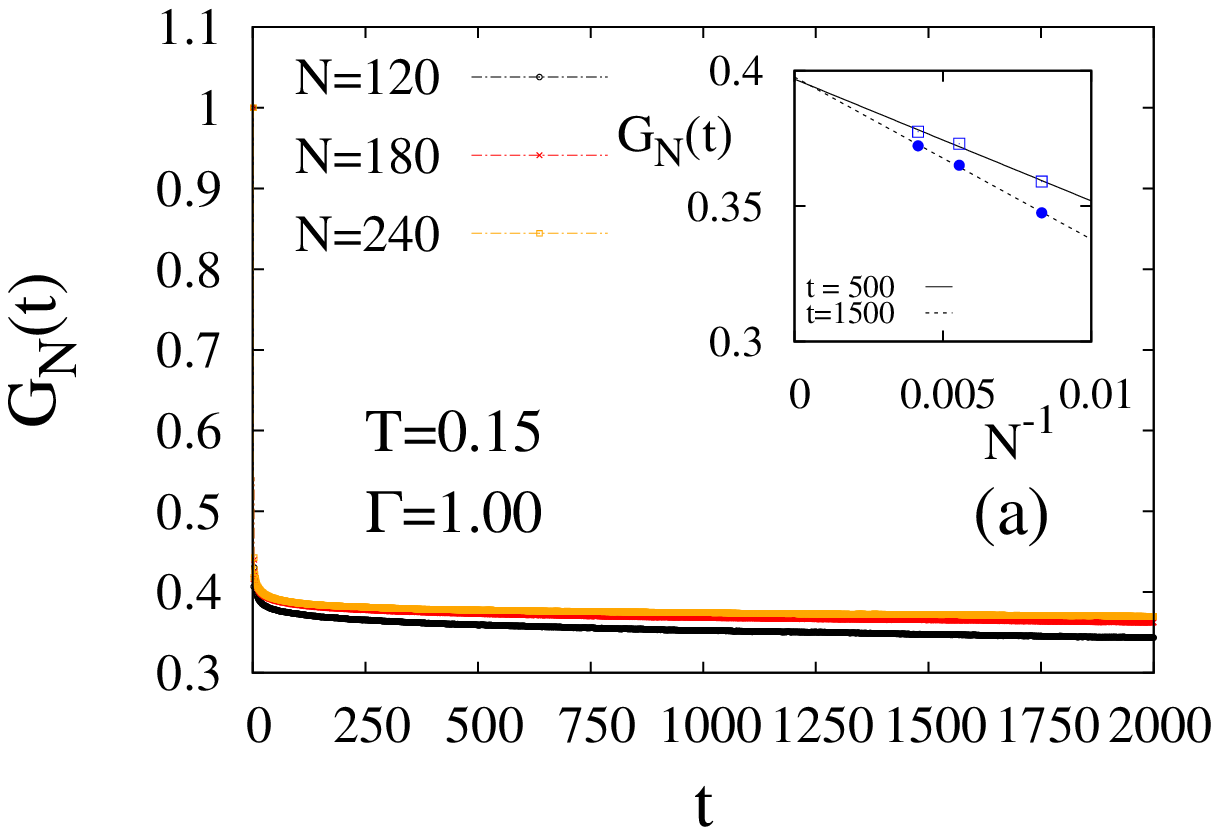}
\includegraphics[width=6.0cm]{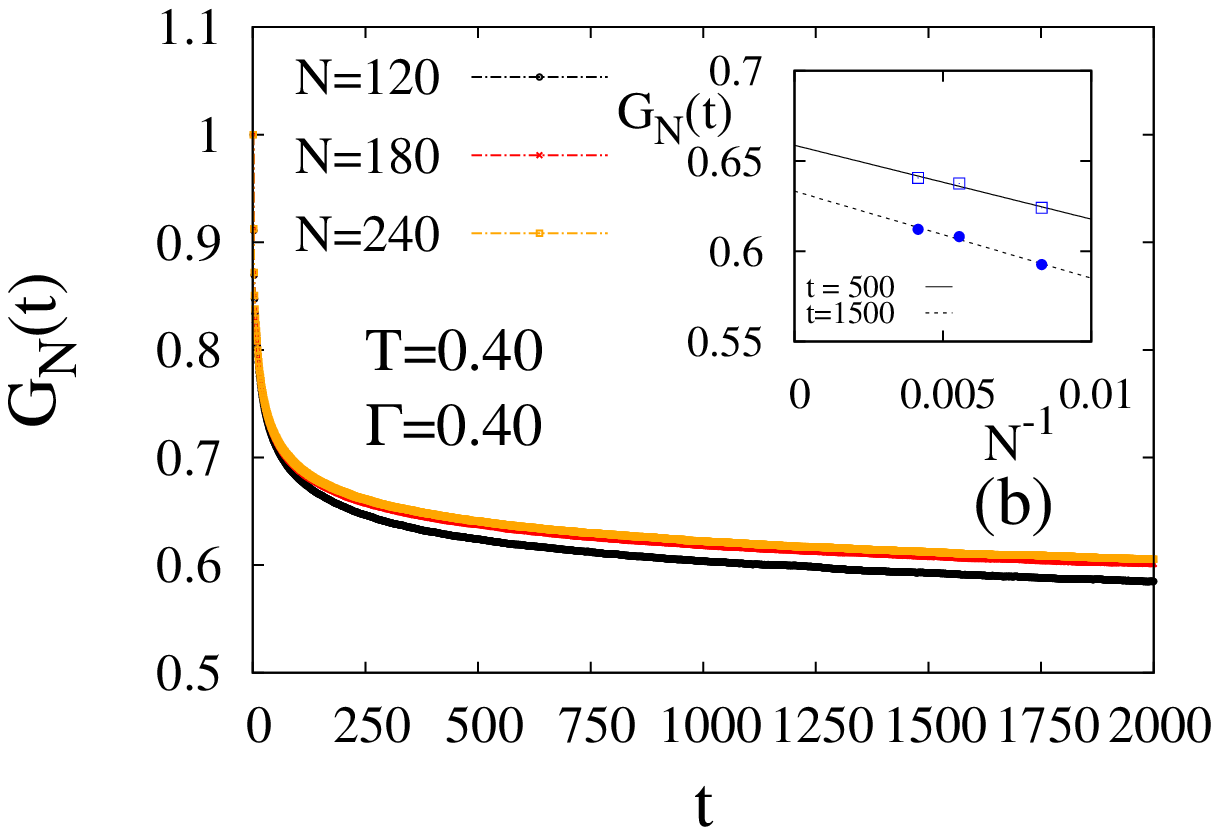}
\includegraphics[width=6.0cm]{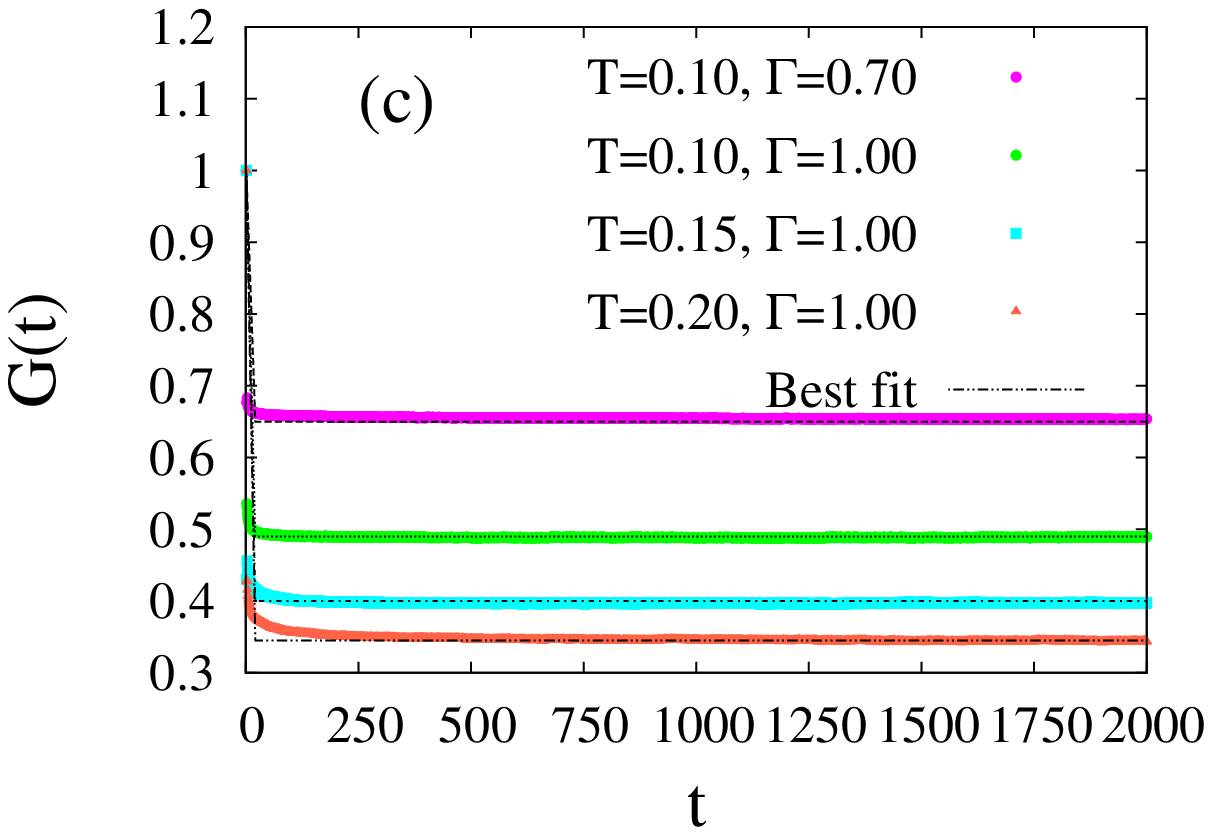}
\includegraphics[width=6.0cm]{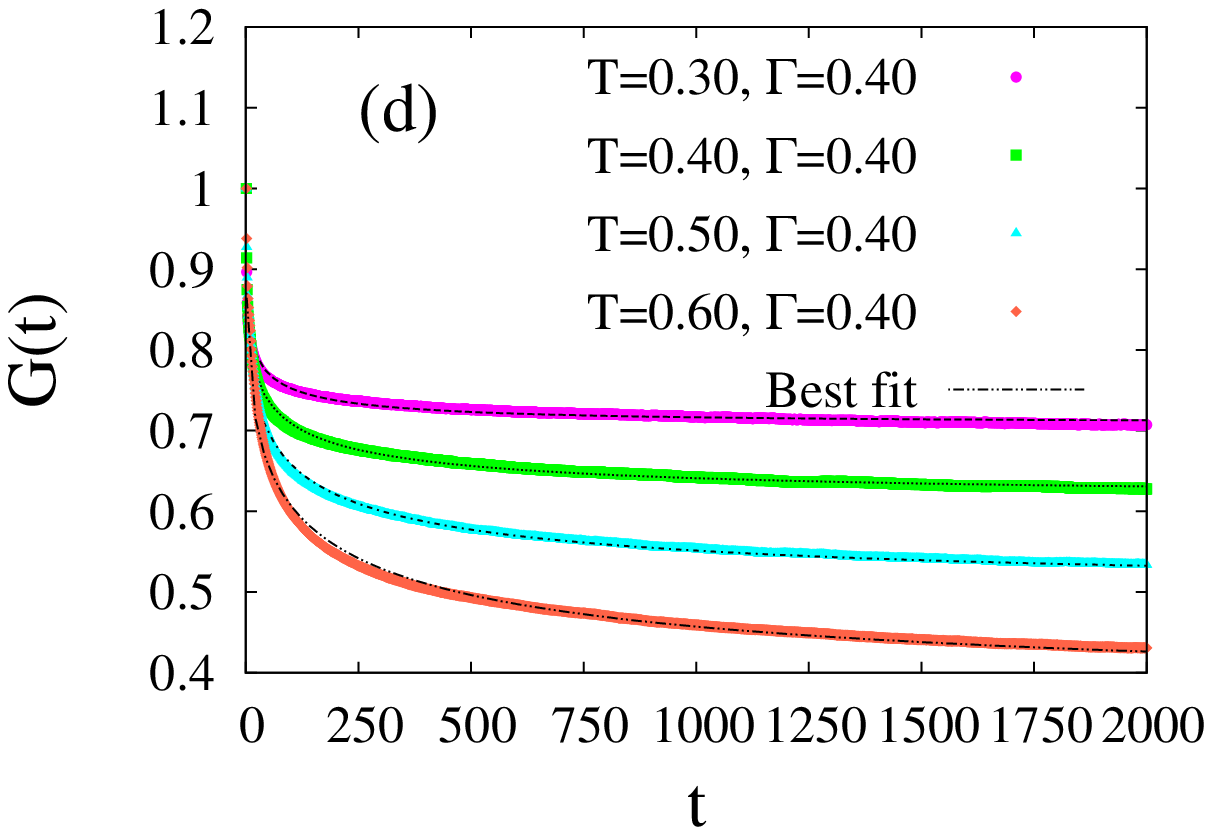}
\end{center}
\caption{ 
The plots of spin autocorrelation function $G_N(t)$ [as defined in Eq.~(\ref{gnt})] with time $t$ for 
 (a) $T = 0.15$ and $\Gamma = 1.00$; (b) $T = 0.40$ and $\Gamma = 0.40$. We take system sizes $N = 120, 180, 240$. 
The extrapolations of $G_N(t)$ with $1/N$ at times $t = 500$ and $1500$, are shown in the intsets. 
Variations of extrapolated autocorrelation $G(t)$ (a) at ($T = 0.10, \Gamma = 0.70$), ($T = 0.10, \Gamma = 1.00$),  
($T = 0.15, \Gamma = 1.00$), and ($T = 0.20, \Gamma = 1.00$); (b) at ($T = 0.30, \Gamma = 0.40$), 
($T = 0.40, \Gamma = 0.40$), ($T = 0.50, \Gamma = 0.50$), and ($T = 0.60, \Gamma = 0.40$). The dotted lines 
show the best-fit [to Eq.~(\ref{auto-corr})] curves associated with these $G(t)$ variations.}
\label{auto_corr_plots}
\end{figure*}

\section{Study of spin autocorrelation dynamics}
We investigate the behavior of the spin autocorrelation in both ergodic and nonergodic spin glass regions~\cite{sudip-jpn}. 
First we take any spin configuration (after the equilibrium) at any particular Monte Carlo step $\tilde{t}$ and then in each 
Monte Carlo step, we calculate the instantaneous overlap of the spin configuration with the spin profile at $t = \tilde{t}$. 
We continue this calculation for an interval of time $\mathbb{T}$. After that we pick the spin configuration at $\mathbb{T} + 1$  
and repeat the same calculation again for an interval of time $\mathbb{T}$. With fixed values of $T$ and $\Gamma$, the 
autocorrelation function $G_N(t)$ for given system size $N$ is define as 
\begin{align}
 G_N(t) = \overline{\Big{\langle} \frac{1}{NM}\sum_{i=1}^N\sum_{n=1}^M\sigma_i^n(t_0)\sigma_i^n(t) \Big{\rangle}}. \label{gnt}
\end{align}
For a given realization of spin-spin interactions, we average $G_N(t)$ over several intervals, which is indicated by $\langle .. \rangle$. 
Again the overhead bar denotes the disorder average. As we compute $G_N(t)$ in the spin glass phase then it should finally 
decay to a finite value. We find the relaxation behaviors of $G_N(t)$ are remarkably different in ergodic and nonergodic spin 
glass regions. We notice that, in the ergodic region, the $G_N(t)$ very quickly achieves its equilibrium value whereas in the 
nonergodic region the relaxation of $G_N(t)$ towards its equilibrium value is much slower than that of the earlier case.

To accomplish the Monte Carlo simulations, we take system sizes $N =120, 180, 240$ and Trotter size $N = 10$. The interval average 
is made over $1000$ intervals where in each interval we take $2000$ Monte Carlo steps. We take $100$ samples for disorder averaging. 
The variation of $G_N(t)$ with $t$ for different system sizes are shown in Fig.~\ref{auto_corr_plots}(a). Here the values of $T = 0.10$ 
and $\Gamma = 1.00$ (belonging to the ergodic region). One can clearly see the quick saturation (almost) of $G_N(t)$ at its equilibrium 
value. In addition to that, one can also notice the system size dependence of $G_N(t)$. To extract the value of $G_N(t)$ in thermodynamic 
limit, we extrapolate $G_N(t)$ with $1/N$ and such extrapolations at $t = 500, 1500$ are shown in the inset of Fig.~\ref{auto_corr_plots}(a). 
We plot the variations of $G_N(t)$ with $t$ for different system sizes in Fig.~\ref{auto_corr_plots}(b) where $T = 0.40$ and $\Gamma = 0.40$ 
(belonging to nonergodic region). Here we again extrapolate  $G_N(t)$ as a function of $1/N$ at $t = 500, 1500$ [see the inset of 
Fig.~\ref{auto_corr_plots}(b)]. In this case we can clearly see the decay of $G_N(t)$ is much slower with respect to the earlier case. 
For given values of $T$ and $\Gamma$, we extract the entire extrapolated autocorrelation curve $G(t)$ for infinite system size. To find 
the relaxation time scale we try to fit $G(t)$ with the function 
\begin{align}
G(t) = G_s + (1 - G_s)\exp\Big[-\Big(\frac{t}{\tau_A}\Big)^{\alpha}\Big]. \label{auto-corr}
\end{align}
The tentative saturation value of $G(t)$ is denoted by $G_s$ and ${\tau}_A$ is the effective relaxation time of the system. Here $\alpha$ 
is the stretched exponent. The extrapolated curves $G(t)$ belong to the ergodic region of the spin glass phase and their associated 
best-fit curves are shown in Fig.~\ref{auto_corr_plots}(c). In the ergodic region, the typical value of relaxation time ${\tau}_A$ is of 
the order of $2$. We find the value of $\alpha$ is very high ($\approx 17 \pm 3$). Similar variations of $G(t)$ (belong to nonregodic spin 
glass region) and their corresponding best-fit lines are shown in Fig.~\ref{auto_corr_plots}(d). In this case, considering $\alpha = 0.31 \pm 0.01$ 
we find reasonably good fittings of $G(t)$ curves. Unlike the ergodic region, in this case we find that the value of ${\tau}_A$ is not 
uniform throughout the nonergodic region. In fact, we notice an increase in the value of $\tau$ as we move towards the deep into the nonergodic spin 
glass region from the line of separation, which separate the ergodic and nonergodic regions. In Table~\ref{chart}, we show the values of 
the $\alpha$ and ${\tau}_A$ obtained form the fittings of the $G(t)$ curves (belonging to both ergodic and nonergodic regions). One 
can notice the change in the value of $\alpha$ as one move from ergodic to nonergodic region. 

\begin{table}[h]
\caption{Best-fit values of $G_s$, $\alpha$, and $\tau_A$ for different 
pairs of $T$ and $\Gamma$, obtained from fitting of $G(t)$ to Eq.~(\ref{auto-corr}).}
\begin{center}
\begin{tabular}{|>{\tiny}c|>{\tiny}c|>{\tiny}c|>{\tiny}c|>{\tiny}c|}
 \hline
 {} & $T = 0.10$, $\Gamma = 1.00$ & $G_s = 0.49$ & $\alpha = 19.75$ & $\tau_A = 1.91$ \\ 
 {Ergodic} & $T = 0.15$, $\Gamma = 1.00$ & $G_s = 0.40$ & $\alpha = 16.64$ & $\tau_A = 1.87$ \\ 
 {(SG)} & $T = 0.20$, $\Gamma = 1.00$ & $G_s = 0.34$ & $\alpha = 14.34$ & $\tau_A = 1.90$ \\
 {} & $T = 0.10$, $\Gamma = 0.70$ & $G_s = 0.65$ & $\alpha = 13.68$ & $\tau_A = 1.86$ \\ \hline
 {} & $T = 0.30$, $\Gamma = 0.40$ & $G_s = 0.71$ & $\alpha = 0.30$ & $\tau_A = 11.01$ \\ 
 {Nonergodic} & $T = 0.40$, $\Gamma = 0.40$ & $G_s = 0.62$ & $\alpha = 0.30$ & $\tau_A =  28.71$ \\ 
 {(SG)} & $T = 0.50$, $\Gamma = 0.40$ & $G_s = 0.51$ & $\alpha = 0.32$ & $\tau_A =  57.20$ \\ 
 {} & $T = 0.60$, $\Gamma = 0.40$ & $G_s = 0.38$ & $\alpha = 0.31$ & $\tau_A =  98.43$ \\ \hline

 \hline
\end{tabular}
\end{center}
\label{chart}
\end{table}

\section{Summary and Discussion}
To extract the nature of the spin glass order parameter distribution (at finite temperature), we perform Monte Carlo simulations 
with system sizes $N = 100, 120, 180, 240$ and Trotter size $M = 15$. Such study for the zero temperature is made through the 
exact diagonalization method with $N = 10, 12, 16, 20$. We should mention that the Monte Carlo results remain fairly 
unaltered when we varied the $M$ with $N$ to keep the $M/N^{z/d}$. Here $z$ is the dynamical exponent and $d$ is the effective 
dimension of the system. We find that in low-temperature (high-transverse-field) spin glass region, the width of the order 
parameter distribution tends to zero for infinite system size. In such region the tail (extended up to zero value of order parameter) 
of the distribution function vanishes in thermodynamic limit. These observations suggest that, with the help of quantum fluctuation 
the system regains its ergodicity in the low-temperature (high-transverse-field) spin glass phase. In contrast to this, we also 
find high-temperature (low-transverse-field) spin glass region where the width as well as the tail of the distribution 
function do not vanish even in thermodynamic limit, which essentially indicates the nonergodic behavior of the system in such 
region of spin glass phase. The possible line which separates the ergodic and nonergodic regions of the quantum spin glass phase, 
is originated from the point ($T = 0.0$ and $\Gamma = 0.0$) and intersects the phase boundary at the quantum-classical-crossover 
point ($T \simeq 0.49$, $\Gamma \simeq 1.31$) [see Fig.~\ref{EE_NEE_phase_diagram}(a)]. 

To study the effect of the quantum-fluctuation-induced ergodicity in the annealing dynamics, we examine the variation 
of the annealing time $\tau$ with the system size. During the course of the annealing,  we tune the temperature and transverse 
field following the schedules $T(t) = T_0(1 - \frac{t}{\tau})$ and ${\Gamma}(t) = {\Gamma}_0(1 - \frac{t}{\tau})$ respectively. 
Here the values of $T_0$ and ${\Gamma}_0$ belong to the paramagnetic phase of the system. 
To avoid the singularities in the Suzuki-Trotter-Hamiltonian at $T = 0$ and $\Gamma = 0$, we assign a nonzero (but very small) 
value for both $T$ and $\Gamma$ even at the end of the annealing schedules. We calculate the required annealing time $\tau$ to 
reach a very low energy state, which is essentially very close the actual ground state of the system. We observe that the 
annealing time $\tau$ becomes clearly system size independent, when the annealing paths go through the ergodic region 
of the spin glass phase. On the other hand, we find $\tau$ becomes strongly system size dependent, when annealing is carried out 
by the paths which pass through the nonergodic region of the spin glass phase [see Fig.~\ref{EE_NEE_phase_diagram}(b)].

We accomplish another finite temperature dynamical study, which again helps to distinguish between the ergodic and nonergodic 
regions in the spin glass phase. Through the Monte Carlo simulations, for a given values of $T$ and $\Gamma$, we study the 
temporal behavior of the average spin autocorrelation $G_N(t)$. For this study we consider system sizes $N = 120, 180, 240$ 
and Trotter size $M = 10$. For each pair of $T$ and $\Gamma$, through the finite-size scaling analysis, we obtain the 
infinite-system-size autocorrelation curve $G(t)$. We attempt to fit $G(t)$ with the stretched exponential functions in Eq.~(\ref{auto-corr}). 
From such fits we find, in the quantum-fluctuation-dominated ergodic spin glass region, the $G(t)$ very quickly relaxes towards 
its equilibrium value within the effective relaxation time ${\tau}_A \sim 2$. We also find the value of the stretched exponent 
$\alpha$ is of the order of $10$, which possibly indicates the fits are not very satisfactory. On the the other hand, in 
the classical-fluctuation-dominated nonergodic region, we find decent fitting of $G(t)$ curves. The obtained values of the effective 
relaxation times are very large compared to the value of ${\tau}_A$ in ergodic region. In this case, we also get $\alpha = 0.31 \pm 0.01$.

Both our static and dynamic studies indicate the regain of ergodicity in the spin glass phase with the aid of quantum-fluctuation. 
In low-temperatures, using the quantum kinetic energy, the system can tunnel through the high (but narrow) free energy barrier. 
As a result of that one would expect the restoration of replica symmetry in the low-temperature (high-transverse field) spin glass phase.  
The effect of the quantum-fluctuation induced ergodicity in the spin glass phase, is also reflected in the dynamical behavior 
of the system, like fast relaxation of spin autocorrelation. The phenomena of quantum tunneling across the macroscopically high 
free energy barriers is not only responsible for making annealing time $\tau$ system-size-independent in the ergodic region of 
the spin glass phase but it is also the origin of the success of the quantum annealing. This ingenious idea of quantum tunneling 
in the free energy landscape of spin glass systems, proposed by Ray et al. 1989, initially faced severe 
criticisms (see e.g.,~\cite{sudip-Goldschmidt}) though eventually their idea is getting
appreciated in the context of quantum annealing  researches. We put an Appendix-A,  highlighting a few such
appreciative sentences  (on Ray et al., 1989) from a chosen set of
20 papers published in the last 5 years. We also give a list of ten top-cited 
publications associated with the quantum annealing studies in Appendix-B.

\newpage\onecolumngrid

\vskip 2cm 

\section*{Appendix-A: Some recent comments on the paper by Ray, Chakrabarti, \& Chakrabarti (1989)}
\vskip 0.5cm 

\noindent The seminal paper by Ray, Chakrabarti, \& Chakrabarti, titled `Sherrington-Kirkpatrick model in a transverse field: 
Absence of replica symmetry breaking due to quantum fluctuations', Physical Review B, vol. 39, p 11828 (1989), first indicated 
the possible benefit of quantum tunneling in the search for the ground state(s) of spin glasses. Although the paper has received modest 
citation (more than 150, Google Scholar), until 2000, most of the citations had been severely critical. It appears, the novelty of their idea 
is gradually being appreciated. We give below some selected sentences from a few sample papers, published in last five years, highlighting their idea 
and giving a telltale version of the the present scenario: 

\begin{enumerate}
\item Boixo et al. (Univ. Southern California, ETH Zurich, ...) in their \textbf{Nature Physics} 
\href{https://www.nature.com/articles/nphys2900}{(2014, Vol. 10, p. 218)} says ``\textit{The phenomenon of \textbf{quantum tunneling} suggests that it \textbf{\textit{can be more efficient to explore the state space quantum mechanically}} in a quantum annealer} [Ray, Chakrabarti \& Chakrabarti Physical Review B (1989); Finnila et al., Chemical Physics Letters (1994); Kadowaki \& Nishimori, Physical Review E (1998)].'' 

\item Cohen \& Tamir (Tel Aviv \& Bar-Ilan Univs.) in their \textbf{International Journal of Quantum Information} 
\href{https://www.worldscientific.com/doi/abs/10.1142/S0219749914300022}{(2014, Vol. 12, art. 1430002)} says  ``\textit{\textbf{Quantum annealing was first discussed by Ray et al. in 1989}} [Ray, Chakrabarti \& Chakrabarti, Physical Review (1989)].". 

\item  Silevitch, Rosenbaum \& Aeppli (Univ. Chicago, Caltech, Swiss Fed. Inst. Tech., etc) in their \textbf{European Physical Journal Special Topics} 
\href{https://epjst.epj.org/articles/epjst/abs/2015/01/epjst2241004/epjst2241004.html}{(2014, vol. 224, p. 25)} say ``\textit{A \textbf{quantum computer has the potential to exploit effects such as entanglement and tunneling} and that appear on the atomic and molecular size scales \textbf{to solve such problems dramatically faster than conventional computers}} [Ray, Chakrabarti \& Chakrabarti, Physical review B (1989); Farhi et al., Science (2001); Santoro et al, Science (2002), Das \& Chakrabarti, Reviews of Modern Physics (2008); Johnson et al., Nature (2011)].". 

\item  Heim et al. (ETH \& Google, Zurich) in \textbf{Science} \href{https://science.sciencemag.org/content/348/6231/215.abstract}{(2015, vol. 348, p. 215)}  say ``\textit{\textbf{Quantum annealing}} [Ray, Chakrabarti \& Chakrabarti, Physical Review B (1989); Finnila et al., Chemical Physics Letters (1994); Kadowaki \& Nishimori, Physical Review E (1998); Farhi et al., Science (2001); Das \& Chakrabarti, Reviews of Modern Physics (2008)] \textit{\textbf{uses quantum tunneling instead of thermal excitations to escape from local minima}, which can be advantageous in systems with tall but narrow barriers, which are easier to tunnel through than to thermally climb over.}".

\item Mandra,  Guerreschi, and Aspuru-Guzik  (Dept. Chem., Harvard Univ. ) in their \textbf{Physical Review A} 
\href{https://journals.aps.org/pra/abstract/10.1103/PhysRevA.92.062320}{(2015, vol. 92, p. 062320)}  begin with the introductory sentence ``\textit{In 2001, Farhi et al.} [Science (2001)] \textit{proposed a new paradigm to carry out \textbf{quantum computation ... that builds on previous results developed} by the statistical \& chemical physics communities \textbf{in the context of quantum annealing techniques}} [Ray, Chakrabarti \& Chakrabarti, Physical Review B (1989); Kadowaki \& Nishimori, Physical Review E (1998); Finnila et al., Chemical Physics Letters (1994); Lee \& Berne, Journal of Physical Chemistry A (2000)].".

\item Boixo et al. (Google \& NASA Ames, California; Michigan State Univ., Michigan; D-Wave Systems \& Simon Fraser Univ., British Columbia;  \& acknowledging discussions with Farhi, Leggett, et al.) in their \textbf{Nature Communications} 
\href{https://www.nature.com/articles/ncomms10327}{(2016, vol. 7, art. 10327)} start the paper with the sentence 
``\textit{\textbf{Quantum annealing}} [Finnila et al. Chemical Physics Letters (1994); Kadowaki \& Nishimori, Physical Review E (1998); Farhi et al., arXiv (2002); Brooke et al., Science (1999); Santoro et al., Science (2002)] \textit{\textbf{is a technique inspired by classical simulated annealing}} [Ray, Chakrabarti \& Chakrabarti, Physical Review B (1989)] \textit{\textbf{that aims to take advantage of quantum tunnelling}.}".

\item Wang, Chen \& Jonckheere  (Dept. Electr. Engg., Univ. Southern California) begin their \textbf{Scientific Reports} 
\href{https://www.nature.com/articles/srep25797}{(2016, vol. 6, art. 25797)} by saying 
``\textit{\textbf{Quantum annealing ... is a generic way to efficiently get close-to-optimum solutions in many NP-hard optimization problems ...  is believed to utilize quantum tunneling instead of thermal hopping to more efficiently search for the optimum solution} in the Hilbert space of a quantum annealing device such as the D-Wave} [Ray, Chakrabarti \& Chakrabarti, Physical Review B (1989); Kadowaki \& Nishimori, Physical Review E (1998)].".

\item Matsuura et al. (Niels Bohr Inst.; Yukawa Inst.; Tokyo Inst. Tech.; Univ. S. California) in their \textbf{Physical Review Letters} 
\href{https://journals.aps.org/prl/abstract/10.1103/PhysRevLett.116.220501}{(2016, vol. 116, p. 220501)} introduce by saying ``\textit{\textbf{Quantum annealing}, a quantum algorithm to solve optimization problems} [Kadowaki \& Nishimori, Physical Review E (1998); Ray, Chakrabarti \& Chakrabarti, Physical Review B (1989); Brooke et al., Science (1999); Brooke et al., Nature (2001); Santoro et al., Science (2002); Kaminsky et al., Quantum Computing (Springer, 2004)] \textit{that is a special case of universal adiabatic quantum computing, \textbf{has garnered a great deal of recent attention as it provides an accessible path to large-scale, albeit nonuniversal, quantum computation using present-day technology.}}".  

\item Yao et al. (Depts. Physics, Univ. California, Berkeley, Harvard Univ., Cambridge, Stanford Univ., California) in their \textbf{arXiv} 
\href{https://arxiv.org/abs/1607.01801}{(2016, no. 1607.01801)}, discussed, both theoretically and experimentally, the fast scrambling or thermal and localized regions of the transverse Ising SK models, following [Ray, Chakrabarti \& Chakrabarti, Physical Review B (1989)] and compared their results (see their Fig. 2 caption) with those reported in [Mukherjee, Rajak \& Chakrabarti, Physical Review E (2015)].

\item Muthukrishnan, Albash \& Lidar (Depts. Physics, Chemistry, Electrical Engineering, ..., Univ. Southern California) write in the Introduction of their \textbf{Physical Review X} \href{https://journals.aps.org/prx/abstract/10.1103/PhysRevX.6.031010}{(2016, vol. 6, p. 031010)}, ``\textit{\textbf{It is often stated that quantum annealing}} [Ray, Chakrabarti \& Chakrabarti, Physical Review B (1989); Finnila et al. Chemical Physics Letters (1994); Kadowaki \& Nishimori, Physical Review E (1998); Farhi et al., Science (2001); Das \& Chakrabarti, Reviews of Modern Physics (2008)] \textit{\textbf{uses tunneling instead of thermal excitations to escape from local minima, which can be advantageous in systems with tall but thin barriers that are easier to tunnel through than to thermally climb over}} [Heim et al., Science (2015); Das \& Chakrabarti, Reviews of Modern Physics (2008), Suzuki, Inoue \& Chakrabarti, Quantum Ising Phases \& Transitions, Springer (2013)]. ... \textit{We demonstrate that the role of 
tunneling is significantly more subtle ...}".

\item Chancellor et al. (Depts. Phys. \& Engg., Univs. Durham, Oxford, London) in the introduction of their \textbf{Scientific Reports} 
\href{https://www.nature.com/articles/srep37107}{(2016, vol. 6, art. 37107)} say 
``\textit{\textbf{There have been many promising advances in quantum annealing, since the idea that quantum fluctuations could help explore rough energy landscapes}} [Ray, Chakrabarti \& Chakrabarti, Physical Review B (1989)], \textit{through the algorithm first being explicitly proposed} [Finnila et al. Chemical Physics Letters (1994)], \textit{further refined} [Kadowaki \& Nishimori, Physical Review E (1998)], \textit{and the basic concepts demonstrated experimentally in a condensed matter system} [Brooke et al., Science (1999)]. ... \text{For an overview ...  see} Das \& Chakrabarti, Reviews of Modern Physics (2008).".

\item Ohzeki (Tohoku University) start his \textbf{Scientific Reports} \href{https://www.nature.com/articles/srep41186}{(2017, vol. 7, art. 41186)} paper with ``\textit{Quantum annealing (QA)... was originally proposed as a numerical computational algorithm} [Kadowaki \& Nishimori, Physical Review E (1998)] \textit{inspired by simulated annealing} [Kirkpatrick, 
Gelatt \& Vecchi, Science (1983)] , \textit{and the exchange Monte Carlo simulation} [Hukushima \& Nemeto Journal of the Physical Society of Japan (1996)]. \textit{\textbf{In QA, the quantum tunneling effect efficiently finds the ground state even in the many-valley structure of the energy landscape therein}} [Ray, Chakrabarti \& Chakrabarti, Physical Review B (1989), Apolloni, Carvalho \& de Falco, Stochastic Process \& their Applications (1989), Das \& Chakrabarti, Reviews of Modern Physics (2008)].". 

\item Azinovic et al. (ETH Zurich; RIKEN, Wako-shi;  Microsoft Research, Redmond; etc.) in their \textbf{SciPost Physics} 
\href{https://scipost.org/submissions/1607.03329v6/}{(2017, vol. 2, art. 013)} says 
``\textit{\textbf{While Simulated Annealing makes use of thermal excitations to escape local minima, quantum annealing}} [Ray, Chakrabarti \& Chakrabarti, Physical Review B (1989); Finniela et al., Chemical Physics Letters (1994), Kadowaki \& Nishimori, Physical Review E (1998); Farhi et al., Science (2001); Das \& Chakrabarti, Reviews of Modern Physics (2008)] \textit{\textbf{uses quantum fluctuations to find the ground state of a system.}}". 

\item Zhang et al. (Stanford Univ., California; Cray, Seattle; Universidad Complutense, Madrid; Univ. Southern California, Los Angeles) in their \textbf{Scientific Reports} \href{https://www.nature.com/articles/s41598-017-01096-6}{(2017, vol. 7, art. 1044)} says in the introduction ``\textit{Quantum annealers} [Kadowaki \& Nishimori, Physical Review E (1998); Farhi et al., Science (2001)] \textit{provide a unique approach to finding the ground-states of discrete optimization problems, utilizing gradually decreasing quantum fluctuations to traverse barriers in the energy landscape in search of global optima, a mechanism commonly believed to have no classical counterpart} [Kadowaki \& Nishimori, Physical Review E (1998); Farhi et al., Science (2001); Finnila et al., Chemical Physics Letters (1994); Brooke et al., Science (1999); Santoro et al., Science (2002); Das \& Chakrabarti, Reviews of Modern Physics (2008); Ray, Chakrabarti \& Chakrabarti, Physical Review B (1989)].".

\item Bottarelli et al. (Univ. Verona, Verona), in their \textbf{FOCUS paper in Soft Computing} 
\href{https://link.springer.com/article/10.1007/s00500-018-3034-z}{(2018, vol. 18/218, art. 29/01/18)} mentions, 
while discussing in the section on 
Quantum Annealing (QA) \& D-Wave quantum annealers/computers, ``\textit{\textbf{The advantage of QA is} the dependency of the tunneling probability both on the height and the width of the potential barrier, which gives it \textbf{the ability to move in an energy landscape where local minima are separated by tall barriers, provided that they are narrow enough}} [Ray, Chakrabarti \& Chakrabarti, Physical Review B (1989)].". 

\item Albash \& Lidar (Univ. Southern California) in their review paper \textbf{Reviews of Modern Physics} 
\href{https://journals.aps.org/rmp/abstract/10.1103/RevModPhys.90.015002}{(2018, vol. 90, p. 015002)} 
note that the exponential run time 
problem in classical annealing comes from ``\textit{... energy barriers in the classical cost that scale with problem size to foil single-spin- update Simulated Annealing (SA). \textbf{This agrees with the intuition that a Stoquastic Adiaabatic Quantum Comuptation advantage over SA is associated with tall and thin barriers}} [Ray, Chakrabarti \& Chakrabarti, Physical Review B (1989); Das \& Chakrabarti, Reviews of Modern Physics (2008)].". 

\item Baldassi \& Zechchina (Bocconi Inst., Milan \& ICTP, Trieste) start their paper \textbf{Proceedings of the National Academy of Science} 
\href{https://www.pnas.org/content/115/7/1457}{(2018, vol. 115, p. 1457)} 
with the sentence ``\textit{\textbf{Quantum annealing aims at finding low-energy configurations} of nonconvex optimization problems \textbf{by a controlled quantum adiabatic evolution}, where a time-dependent many-body quantum system which encodes for the optimization problem evolves toward its ground states so as \textbf{to escape local minima through multiple tunneling events}} [Ray, Chakrabarti \& Chakrabarti, Physical Review B (1989); Finnila et al., Chemical Physics Letters (1994); Kadwaki \& Nishimori, Physical Review E (1998); Farhi et al., Science (2001); Das \& Chakrabarti, Reviews of Modern Physics (2008)].". 

\item Mishra, Albash \& Lidar (Depts. Physics, Chemistry, Electrical Engineering, Univ. Southern California) begin their paper \textbf{Nature Communications} \href{https://www.nature.com/articles/s41467-018-05239-9}{(2018, vol. 9, art. 2917)} with the sentence ``\textit{Quantum annealing} [Apolloni, Carvalho \& de Falco, Sotcastic Processes \& Applications (1989); Apolloni, Cesa-Bianchi \& de Falco, in Stochastic Process, Physics \& Geometry, World Scientific (1990); Ray, Chakrabarti \& Chakrabarti, Physical Review B (1989); Somoraji, Journal of Physical Chemistry (1991); Amara, Hsu \& Straub (1993), Journal of Physical Chemistry (1993); Finnila et al., Chemical Physics Letters (1994); Kadwaki \& Nishimori, Physical Review E (1998); Das \& Chakrabarti, Reviews of Modern Physics (2008)], \textit{also known as 
the quantum adiabatic algorithm} [Farhi et al. arXiv (2000); Farhi et al., Science (2001)] \textit{or adiabatic quantum optimization} [Smelyanski, Toussaint \& Timukin, arXiv (2001); Reichardt, in Proceedings of the ACM Symposium on Theory of Computing: ACM-36 (2004)] \textit{is a heuristic quantum algorithm for solving combinatorial optimization problems.}". 

\item Jiang et al. (Dept. Computer Science, Purdue Univ., Quantum Computing Institute, Oak Ridge National Laboratory) reporting on their quantum annealing framework for prime number factorization in \textbf{Scientific Reports} 
\href{https://www.nature.com/articles/s41598-018-36058-z}{(2018, vol. 8, art. 17667)} write ``\textit{In this contribution, we introduce a new procedure for solving the integer factorization problem using quantum annealing} [Kadwaki \& Nishimori, Physical Review E (1998); Das \& Chakrabarti, Reviews of Modern Physics (2008)] \textit{which utilizes adiabatic quantum computation. ... Quantum Annealing was introduced} [Kadwaki \& Nishimori, Physical Review E (1998)] \textit{to solve optimization problems using quantum fluctuations to transit to the ground state, compared to simulated annealing which uses thermal fluctuations to get to the global minimum. \textbf{Quantum fluctuations such as quantum tunneling}} [Ray, Chakrabarti \& Chakrabarti, Physical Review B (1989)] \textit{\textbf{provide ways of transitions between states. The transverse 
field controls the rate of the 
transition, as the role of temperature played in simulated annealing.}}".  

\item Sato et al. (School of Sc. \& Engg., Saitama Univ.; Fujitsu Laroratories; Japan Science \& Technology) write in the Introduction of their paper \textbf{Physical Review E} \href{https://journals.aps.org/pre/abstract/10.1103/PhysRevE.99.042106}{(2019, vol. 99, p. 042106)} ``\textit{\textbf{There are two famous annealing concepts}} [Kirkpatrick, Gelatt Jr. \& Vecchi, Science (1983); Ray, Chakrabarti \& Chakrabarti, Physical Review B (1989)]: \textit{\textbf{one is the simulated annealing method in which the temperature of the system is controlled to search the global minimum; another is the quantum annealing method which uses quantum effects.}}". 
\end{enumerate}

\vskip 1cm 

\rightline {\small{[The highlights in the text of each citation is by the author.]}}

\newpage 

\section*{Appendix-B: Ten top-cited publications from Google Scholar with `Quantum Annealing' in the title}  
\begin{enumerate}
\item \textit{Quantum annealing in the transverse Ising model}, T Kadowaki and H Nishimori,  Physical Review E, \textbf{58}, 5355 (1998) [803]. 

\item \textit{Quantum annealing with manufactured spins}, M. W. Johnson, M. H. S. Amin, S. Gildert, T. Lanting, F. Hamze, N. Dickson,
R. Harris, A. J. Berkley, J. Johansson, P. Bunyk, E. M. Chapple, C. Enderud,
J. P. Hilton, K. Karimi, E. Ladizinsky, N. Ladizinsky, T. Oh, I. Perminov, C. Rich,
M. C. Thom, E. Tolkacheva, C. J. S. Truncik, S. Uchaikin, J. Wang, B. Wilson, 
G. Rose, Nature, \textbf{473}, 194 (2011) [784].  

\item \textit{Evidence for quantum annealing with more than one hundred qubits}, S. Boixo, T. F. Rønnow, S. V. Isakov, Z. Wang, D. Wecker, 
D. A. Lidar, J. M. Martinis, M. Troyer, Nature Physics, \textbf{10}, 218 (2014) [460]. 

\item \textit{Theory of quantum annealing of an Ising spin glass}, 
G. E. Santoro, R. Martoňák, E. Tosatti, R. Car, Science, \textbf{295}, 2427 (2002) [442].

\item \textit{Quantum annealing: A new method for minimizing multidimensional functions}, 
A. B. Finnila, M. A. Gomez, C. Sebenik, C. Stenson,  J. D. Doll, Chemical Physics Letters, \textbf{219}, 343 (1994) [400].  
  
\item \textit{Colloquium: Quantum annealing and analog quantum computation}, A. Das and B. K. Chakrabarti, Reviews of Modern Physics, 
 \textbf{80}, 1061 (2008) [385]. 
 
\item \textit{Quantum annealing of a disordered magnet},  J. Brooke, D. Bitko,  T. F. Rosenbaum, G. Aeppli, Science, \textbf{284}, 779 (1999) [380].   

\item \textit{Optimization using quantum mechanics: quantum annealing through adiabatic evolution}, G. E. Santoro and E. Tosatti,  
Journal of Physics A: Mathematical and General, \textbf{39}, R393 (2006) [238]. 

\item \textit{Entanglement in a quantum annealing processor}, 
T. Lanting, A. J. Przybysz, A. Yu. Smirnov, F. M. Spedalieri, M. H. Amin, A. J. Berkley, R. Harris, F. Altomare, S. Boixo, P. Bunyk, N. Dickson, 
C. Enderud, J. P. Hilton, E. Hoskinson, M. W. Johnson, E. Ladizinsky, N. Ladizinsky, R. Neufeld, T. Oh, I. Perminov, C. Rich, M. C. Thom, 
E. Tolkacheva, S. Uchaikin, A.B. Wilson, G. Rose, Physical Review X, \textbf{4}, 021041 (2014) [212]. 

\item \textit{Quantum annealing and related optimization methods}, Eds. A. Das and B. K. Chakrabarti, Springer, Heidelberg (2005) [203]. 
\end{enumerate}

\vskip 1cm 

\small{[Numbers in `[~]' give the corresponding number of citations as on 22nd April, 2019, when the list had been compiled.]}  
\end{document}